\documentclass [11pt, letterpaper]{article}
\usepackage{fullpage}
\usepackage{amsmath}
\usepackage{amssymb}
\usepackage{graphicx}
\usepackage{mathrsfs}
\usepackage{wrapfig}
\usepackage{boxedminipage}
\usepackage{epsfig}
\usepackage{setspace}
\usepackage{subfigure}
\usepackage{float}
\usepackage{hyperref}
\usepackage{enumerate}
\newcommand{\reef}[1]{(\ref{#1})}
\begin{document}

\begin{flushright}
\phantom{{\tt arXiv:1202.????}}
\end{flushright}

\bigskip
\bigskip
\bigskip

\begin{center} {\Large \bf Holographic Studies}
  
  \bigskip

{\Large\bf  of }

\bigskip

{\Large\bf     Entanglement Entropy in Superconductors}

\end{center}

\bigskip \bigskip \bigskip \bigskip

\centerline{\bf Tameem Albash, Clifford V. Johnson}

\bigskip
\bigskip
\bigskip

  \centerline{\it Department of Physics and Astronomy }
\centerline{\it University of
Southern California}
\centerline{\it Los Angeles, CA 90089-0484, U.S.A.}

\bigskip

\centerline{\small \tt talbash,  johnson1,  [at] usc.edu}

\bigskip
\bigskip


\begin{abstract} 
\noindent 
We present the results of our studies of the entanglement entropy of a superconducting system described holographically as a fully back--reacted gravity system, with a stable ground state. We use the holographic prescription for the entanglement entropy. We uncover the behavior of the entropy across the superconducting phase transition, showing the reorganization of the degrees of freedom of the system. We exhibit the behaviour of the entanglement entropy from the superconducting transition all the way down to the ground state at $T=0$. In some cases, we also observe a novel  transition in the entanglement entropy  at intermediate temperatures, resulting from the detection of an additional length scale.

\end{abstract}
\newpage \baselineskip=18pt \setcounter{footnote}{0}

\section{Introduction}
%
Given a quantum system, the entanglement entropy  of a subsystem $\cal A$ and its complement $\cal B$ is defined as follows:
\begin{equation}
S_{\mathcal{A}} = - \mathrm{Tr}_{\mathcal{A}} \left( \rho_\mathcal{A} \ln \rho_\mathcal{A} \right)\ ,
\end{equation}
where $\rho_\mathcal{A}$ is the reduced density matrix of $\mathcal{A}$ given by tracing over the degrees of freedom of $\mathcal{B}$,
$\rho_\mathcal{A} = \mathrm{Tr}_{\mathcal{B}}( \rho) $,
where $\rho$ is the density matrix of the system.  

The entanglement entropy is understood as an important probe of physics in various domains, and for systems at strong coupling it is looked upon as a robust tool for keeping track of the degrees of freedom when other traditional probes  (such as an order parameter) might not be available. However, it is often difficult to compute the entanglement entropy in such systems, especially outside 1+1 dimensions. 

Two developments in the field have made the work in this paper possible.  The first is that the entanglement entropy has a  natural geometrical
 definition~\cite{Ryu:2006bv,Ryu:2006ef} (proposed but only partially proven\footnote{See \emph{e.g.,}
refs.~\cite{Solodukhin:2006xv,Hirata:2006jx,Nishioka:2006gr,Headrick:2007km,Solodukhin:2008dh,Casini:2008as,Nishioka:2009un,Casini:2011kv}.}.) in the context of gauge/gravity duals, where a wide class of strongly coupled theories in $d$ dimensions can be defined holographically as dual to a theory of gravity (plus other degrees of freedom) in $d+1$ dimensions, which is in turn ultimately embedded into a 10 dimensional superstring theory or an 11 dimensional supergravity background as a means of ensuring full quantum consistency. Of course, holography is also not fully proven, but there is a large body of evidence for it in numerous examples, starting with the AdS/CFT correspondence and deformations and generalizations thereof~\cite{Maldacena:1997re,Gubser:1998bc,Witten:1998qj,Witten:1998zw}. In this paper we shall assume that holography is robust, and that the holographic formula (reviewed below) for the entanglement entropy is also correct. 

The second development is that  some of the strongly coupled physics of interest, superconducting\footnote{To be precise, the physics breaks a global, not local, symmetry, but it is close to being gauged, in a sense~\cite{Hartnoll:2008kx}. So rather than using the term superfluidity, we will continue with the common usage.} phases that share various features with certain exotic phases of experimentally studied strongly coupled quantum systems, can not only be modelled holographically as an effective model of gravity plus a scalar (see {\it e.g.} ref.~\cite{Hartnoll:2008vx}), not only be embedded consistently  into the parent superstring theory and/or 11D supergravity to get access to the back reactions of the scalar dynamics on the geometry (see {\it e.g.} ref.~\cite{Gauntlett:2009dn}), but can be embedded in a manner that appears to be highly {\it stable}\footnote{Strictly speaking, the stability is typically studied in various truncations of 11D supergravity to lower dimensions.  We mean here full perturbative stability of the ground state in {\it maximal} $\mathcal{N}=8$ supergravity in 4D. There remains the possibility of instabilities arising upon uplift to the full 11D supergravity, non--perturbative instabilities, and parts of the phase diagram being modified at higher temperatures by instabilities of the sort discussed in {\it e.g.} ref.\cite{Gubser:2000ec}. } ({\it i.e.,} refs.~\cite{Bobev:2010ib} and \cite{Cassani:2011fu} have shown that the ground states of an infinite subset of the family of superconductors defined by the embeddings in ref.~\cite{Gauntlett:2009dn} are in fact unstable in {\it maximal} $\mathcal{N}=8$ supergravity in 4D and  it remains to be shown whether any of the others in the family are stable). In other words, there is  a holographic superconductor model (presented recently in ref.~\cite{Bobev:2011rv}) that has a ground state that, thought of as a holographic flow\cite{Gubser:2009qm}
is a stable\footnote{Subject to the caveats in the previous footnote.} non--supersymmetric vacuum of the full theory. This  suggests that  the complete theory of gravity plus all the attendant fields is without pathological physics that might obscure the lessons to be learned from it about strongly coupled phenomena pertinent to the superconductivity.

These two developments come together nicely since to employ the holographic definition of the entanglement entropy in a study of superconductivity, we need the complete (back--reacted--upon and stable) geometry to perform the computation. In this paper we  carry out the study of the entanglement entropy in holographic superconductivity for the first time using these methods, and uncover some very interesting phenomena.

\begin{figure}[h]
\begin{center}
{\includegraphics[width=3.0in]{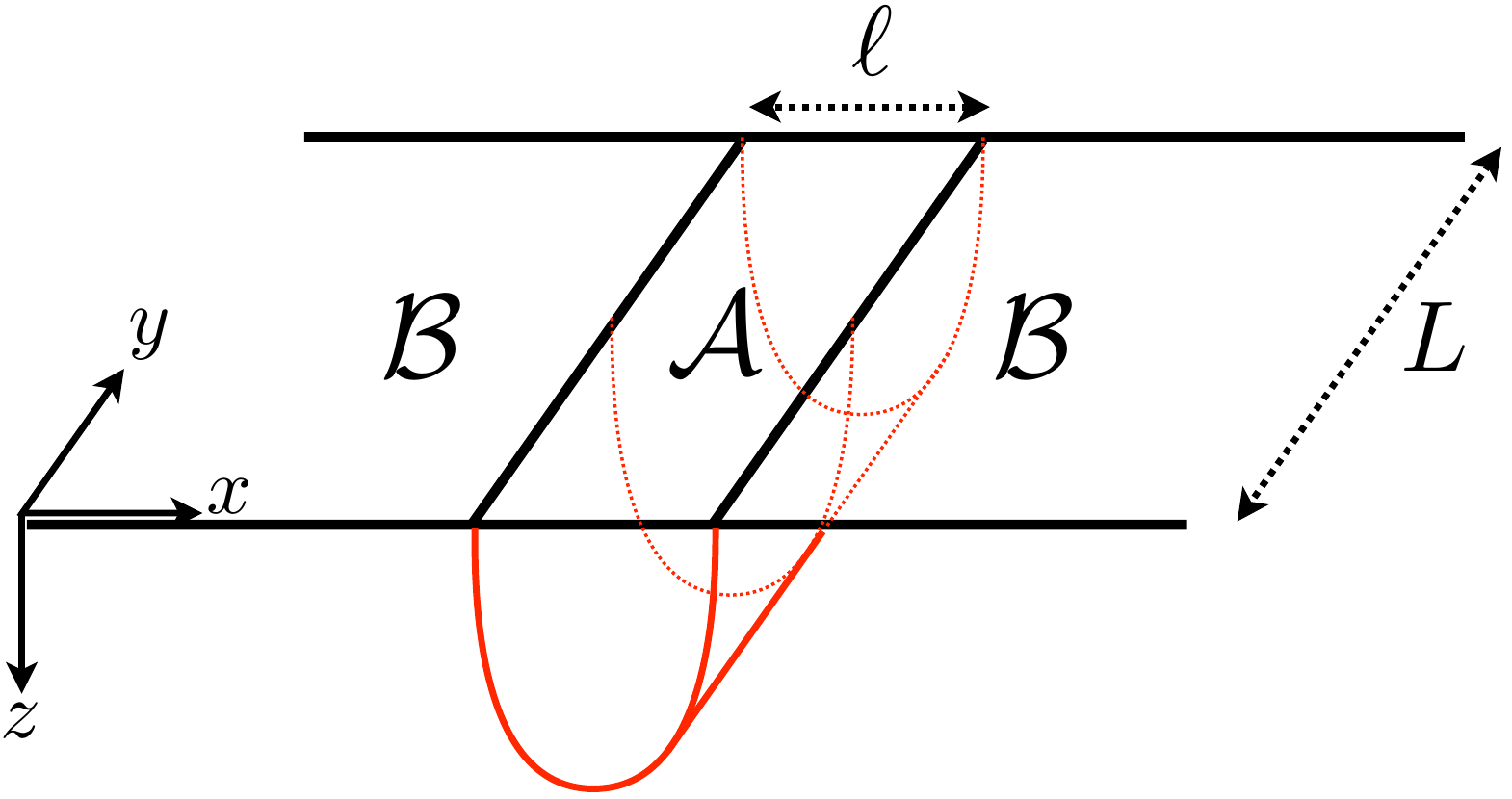}} 
   \caption{\small Diagram of the strip shape we will consider for region $\mathcal{A}$. This is the case of a dual geometry that is asymptotically AdS$_4$, and here,  $z$ denotes the radial direction in AdS$_4$. The quantity $\ell$ sets the size of region $\cal A$, and $L$ is a regulator that is understood to be taken to infinity.}  \label{fig:setup_shapes}
   \end{center}
\end{figure}

The entropy is holographically computed as follows~\cite{Ryu:2006bv,Ryu:2006ef}.
 In an asymptotically Anti--de
Sitter (AdS) geometry, consider a slice at constant AdS radial
coordinate $z = a$. Recall that this defines the dual field theory
(with one dimension fewer) as essentially residing on that slice in
the presence of a UV cutoff set by the position of the slice. Sending
the slice to the AdS boundary at infinity removes the cutoff (see
ref. \cite{Aharony:1999t} for a review).  On our $z=a$ slice, consider
a region $\mathcal{A}$. Now find the minimal--area surface
$\gamma_{\mathcal{A}}$ bounded by the perimeter of $\mathcal{A}$ and
that extends into the bulk of the geometry.  (Figure~\ref{fig:setup_shapes} shows examples of the arrangement we will consider in this paper.)
Then the entanglement
entropy of region $\mathcal{A}$ with $\mathcal{B}$ is given by:
\begin{equation}
\label{eq:holographic_entanglement}
S_{\mathcal{A}} = \frac{ \mathrm{Area}(\gamma_{\mathcal{A}})}{4 G_{\rm N}} \ ,
\end{equation}
where $G_{\rm N}$ is Newton's constant in the dual gravity
theory. 

In the next section, we will review the four dimensional model of gravity plus scalars and a gauge field that was presented in ref.~\cite{Bobev:2011rv}, and review and re--derive the properties of  the solutions we need.

%
\section{Gravity Background}
The Lagrangian of ref.~\cite{Bobev:2011rv} arises as an $SO(3)\times SO(3)$ invariant truncation of four--dimensional  ${\cal N}=8$ gauged supergravity:
\begin{equation}
\frac{\mathcal{L}}{\sqrt{-G} } = \frac{1}{16 \pi G_4} \left( \mathcal{R}  - \frac{1}{4} F_{\mu \nu} F^{\mu \nu} -  2 \partial_\mu \lambda \partial^\mu \lambda -\frac{\sinh^2 \left( 2 \lambda \right)}{2} \left( \partial_\mu \varphi - \frac{g}{2} A_{\mu} \right) \left(  \partial^\mu \varphi  - \frac{g}{2} A^{\mu} \right) - \mathcal{P}  \right) \ ,
\end{equation}
where the potential $\mathcal{P}$ is given by:
\begin{equation} \label{eqt:P}
\mathcal{P} = - g^2 \left( 6 \cosh^4 \lambda - 8 \cosh^2 \lambda \sinh^2 \lambda + \frac{3}{2} \sinh^4 \lambda \right) \ .
\end{equation}
We use a different notation from ref.~\cite{Bobev:2011rv}.  We first  reintroduced the dimensionful constant $8 \pi G_4$  and then  made the following field redefinitions:
\begin{equation}
A_\mu \to \frac{1}{\sqrt{16 \pi G_4}} A_\mu \ , \quad g \to \frac{g}{\sqrt{2}} \ , \quad \mathcal{P} \to \frac{\mathcal{P}}{2} \ .
\end{equation}
Note that the gauge field $A_\mu$ and the (real) scalar fields $\lambda$ and $\varphi$ are dimensionless in this framework. 

The metric and other fields of interest are parameterized as follows:
\begin{equation}
d s^2 = - \frac{R^2}{z^2} f(z) e^{- \chi(z) } dt^2 + \frac{R^2}{z^2} \left( d x_1^2 + d x_2^2 \right) + \frac{R^2}{z^2} \frac{d z^2}{f(z)} \ , \quad A_t = \Psi(z)  \ , \quad \lambda = \lambda(z) \ ,
\end{equation}
and the scalar $\varphi$  will be set to zero. Defining a useful dimensionless coordinate:
\begin{equation}
z = R  \tilde{z} \ ,
\end{equation}
the equations of motion can be reduced to:
\begin{eqnarray}
&&\chi' - 2 \tilde{z} \left( \lambda' \right)^2 - \frac{\tilde{z} e^{\chi} \sinh^2 \left( 2 \lambda \right) \Psi^2 }{8 f^2} = 0 \ , \\
&&\left( \lambda' \right)^2 - \left( \frac{f'}{\tilde{z} f}  \right) + \frac{\tilde{z}^2 e^{\chi} \left( \Psi' \right)^2}{4 f} + \frac{R^2 \mathcal{P}}{2 \tilde{z}^2 f} + \frac{3}{\tilde{z}^2} + \frac{e^{\chi} \sinh^2 \left( 2 \lambda \right) \Psi^2}{16 f^2} = 0 \ , \\
&& \Psi'' + \left( \frac{\chi'}{2} \right) \Psi' - \frac{\sinh^2 \left( 2 \lambda \right) \Psi}{4 \tilde{z}^2 f} = 0 \ , \\
&& \lambda'' + \left( - \frac{\chi'}{2} + \frac{f'}{f} - \frac{2}{\tilde{z}} \right) \lambda' - \frac{R^2}{4 \tilde{z}^2 f} \frac{d \mathcal{P}}{d \lambda} + \frac{ e^{\chi} \sinh \left( 4 \lambda \right) \Psi^2}{16 f^2} = 0 \ .
\end{eqnarray}
The ultraviolet (UV) asymptotic behavior (near the AdS boundary $z = 0$):
\begin{eqnarray}
\lambda(\tilde{z}) &=& \lambda_1 \tilde{z} + \lambda_2 \tilde{z}^2 + \dots \ , \nonumber \\
\chi(\tilde{z}) &=&\chi_0 + \lambda_0^2 \tilde{z}^2 + \dots \ , \nonumber \\
f(\tilde{z}) &=& 1 + \lambda_0^2 \tilde{z}^2 + f_3 \tilde{z}^3 + \dots \ , \nonumber \\
\Psi(\tilde{z}) &=& \Psi_0 + \Psi_1 \tilde{z} + \dots \ .
\end{eqnarray}
Generically we will be at finite temperature, to which there will be associated an event horizon in the geometry. We assume the event horizon occurs at $\tilde{z} = \tilde{z}_H$, and near there the fields have an expansion:
\begin{eqnarray}
\lambda(\tilde{z}) &=& \lambda^{(0)} + \lambda^{(1)} \left( 1 - \frac{\tilde{z}}{\tilde{z}_H} \right) + \dots \ , \nonumber \\ 
\chi(\tilde{z}) &=&  \chi^{(0)} + \chi^{(1)} \left( 1 - \frac{\tilde{z}}{\tilde{z}_H} \right) + \dots \ , \nonumber \\ 
f(\tilde{z}) &=& f^{(1)}  \left( 1 - \frac{\tilde{z}}{\tilde{z}_H} \right) + \dots \ , \nonumber \\ 
\Psi(\tilde{z}) &=& \Psi^{(1)} \left( 1 - \frac{\tilde{z}}{\tilde{z}_H} \right) + \dots \ .
\end{eqnarray}
There are only three independent parameters here, which we choose to be $\lambda^{(0)}, \ \chi^{(0)}, \ \Psi^{(1)}$.  There are three scaling symmetries of the equations of motion given by \cite{Bobev:2011rv}:
\begin{eqnarray}
& t \to \gamma_1^{-1} t \ , \quad \chi \to \chi - 2 \ln \gamma_1 \ , \quad \Psi \to \gamma_1 \Psi \ , &  \label{eqt:scaling1}\\
& t \to \gamma_2^{-1} t \ , \quad z \to \gamma_2^{-1} z \ , \quad R \to \gamma_2^{-1} R \ , &  \label{eqt:scaling2} \\
& x^\mu \to \gamma^{-1} x^\mu \ , \quad f(z) \to f(z) \ , \quad \Psi(z) \to \gamma \Psi(z) \ , \quad \lambda(z) \to \lambda(z) \ , \quad \chi(z) \to \chi(z) \ . & \label{eqt:scaling3}
\end{eqnarray}
Using these scaling symmetries, we can choose any value for the position of the event horizon and the asymptotic value of the field $\chi(z)$.  We choose to  fix $\lambda^{(0)}$ and  $\chi^{(0)}$,  and tune $\Psi^{(1)}$ to either have $\lambda_1$ or $\lambda_2$ be zero. These asymptotic values of the field $\lambda$ in the UV  define the vacuum expectation value (vev) of charged operators in the theory that are either of dimension 1 or 2, and we correspondingly call them ${\cal O}_1$ and ${\cal O}_2$. We will explicitly identify the correctly normalized relationship below. The UV asymptotics of the electric gauge field  component $\Psi(\tilde{z})$ defines a chemical potential and charge density that will be explicitly identified below.

Generically, a solution will have $\chi_0$ non--zero.  To recover pure AdS, we use the scaling symmetry in equation \reef{eqt:scaling1} to shift $\chi(z)$ as:
\begin{equation}
\tilde{\chi}(z)  = \chi(z) - \chi_0 \ ,
\end{equation}
which can be accomplished by rescaling the time coordinate:
\begin{equation}
\tilde{t} = e^{- \chi_0 / 2} t \ ,
\end{equation}
which in turn means:
\begin{equation}
A_{\tilde{t}} =e^{\chi_0/2} A_t \ .
\end{equation}
The temperature of the system is then given by \cite{Bobev:2011rv}:
\begin{equation}
T = \frac{1}{4 \pi R \tilde{z}_H}\frac{e^{-\left(\chi^{(0)} - \chi_{0} \right) /2}}{32} \left( 61 + 36 \cosh\left( 2 \lambda^{(0)} \right) - \cosh \left( 4 \lambda^{(0)} \right) - 8 \tilde{z}_H^2 e^{\chi^{(0)}} \left( \Psi^{(1)}\right)^2 \right)\ ,
\end{equation}
and the chemical potential $\mu$ and charge density $\rho$ go as:
\begin{equation}
\mu = \frac{e^{\chi_0/2}}{\sqrt{16 \pi G_4}} \Psi_0 \  , \quad \rho = - \frac{e^{\chi_0/2}}{R \sqrt{16 \pi G_4}} \Psi_1 \ ,
\end{equation}
and the vevs of the two operators are defined as:
\begin{equation}
\mathcal{O}_1 = \frac{2 \lambda_1}{\sqrt{16 \pi G_4}} \ , \quad \mathcal{O}_2 = \frac{2 \lambda_2}{\sqrt{16 \pi G_4} R} \ .
\end{equation}
Using the scaling symmetry in equation \reef{eqt:scaling3}, the relevant quantities for us scale as:
\begin{equation}
T \to \gamma_3 T \ , \quad \rho \to \gamma_3^2 \rho \ , \quad \mathcal{O}_1 \to \gamma_3 \mathcal{O}_1 \ , \quad \mathcal{O}_2 \to \gamma_3^2 \mathcal{O}_2 \ .
\end{equation}
Therefore, we will use the following  dimensionless quantities to examine the physics:
\begin{equation}
\frac{T}{\sqrt{\rho}} \ , \quad \frac{\mathcal{O}_1}{\sqrt{\rho}} \ , \quad \frac{\mathcal{O}_2}{\rho} \ .
\end{equation}
%
\subsection{High Temperature Phase}
%
At high temperature, the solution is simply the  Reissner--Nordstr\"om AdS solution.  The scalar profile $\lambda(z)$ being zero means that there is no condensate, {\it i.e.,} the operators ${\cal O}_1$ and ${\cal O}_2$ vanish. The solution is given by taking (we restore dimensionful $z$ for now):
\begin{equation}
\lambda(z) = 0 \ , \quad \chi(z) = 0 \ , \quad \Psi(z) = \frac{2 Q R}{z_H} \left( 1 - \frac{z}{z_H} \right) \ , \quad f(z) = 1 + Q^2 \frac{z^4}{z_H^4} - \frac{z^3}{z_H^3} \left( 1 + Q^2 \right) \ .
\end{equation}
So we read off the temperature, chemical potential, and density as:
\begin{equation}
T=\frac{1}{4\pi z_H}(3-Q^2)\ ,\quad \mu = \frac{R}{\sqrt{16 \pi G_4}} \frac{2Q}{z_H}\  , \quad \rho =  \frac{R}{ \sqrt{16 \pi G_4}} \frac{2Q}{z_H^2}\ .
\end{equation}
%
%
\subsection{Intermediate Temperature Phase}
%
At  low enough temperatures, a new type of solution is available  that is a charged black hole with a non--zero charged scalar profile.  We have non--zero $\lambda(z)$ and $\chi(z)$, and  we require either $\lambda_1 = 0$ or $\lambda_2 = 0$, corresponding to having either $\mathcal{O}_2$ or $\mathcal{O}_1$ turned on respectively.   

\begin{figure}[h]
\begin{center}
   \subfigure[$\mathcal{O}_1$]{\includegraphics[width=2.8in]{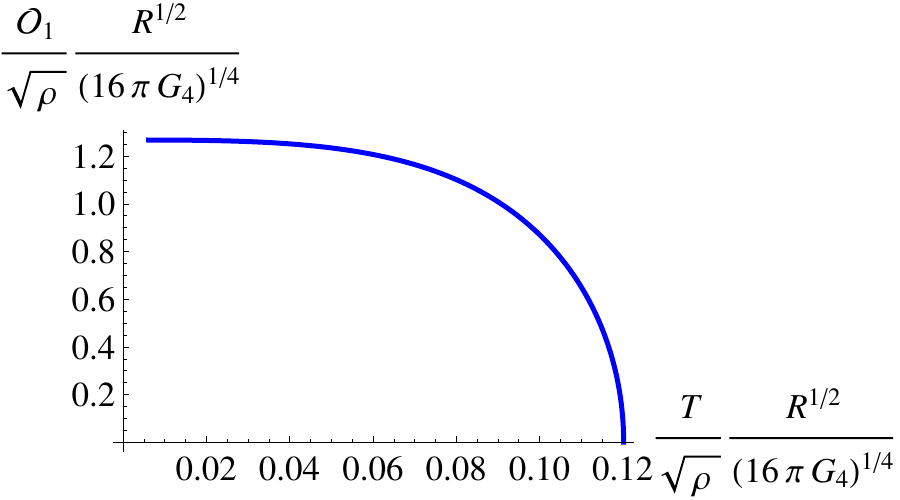}\label{fig:O1_vs_T}} \hspace{0.5cm}
\subfigure[$\mathcal{O}_2$]{\includegraphics[width=2.8in]{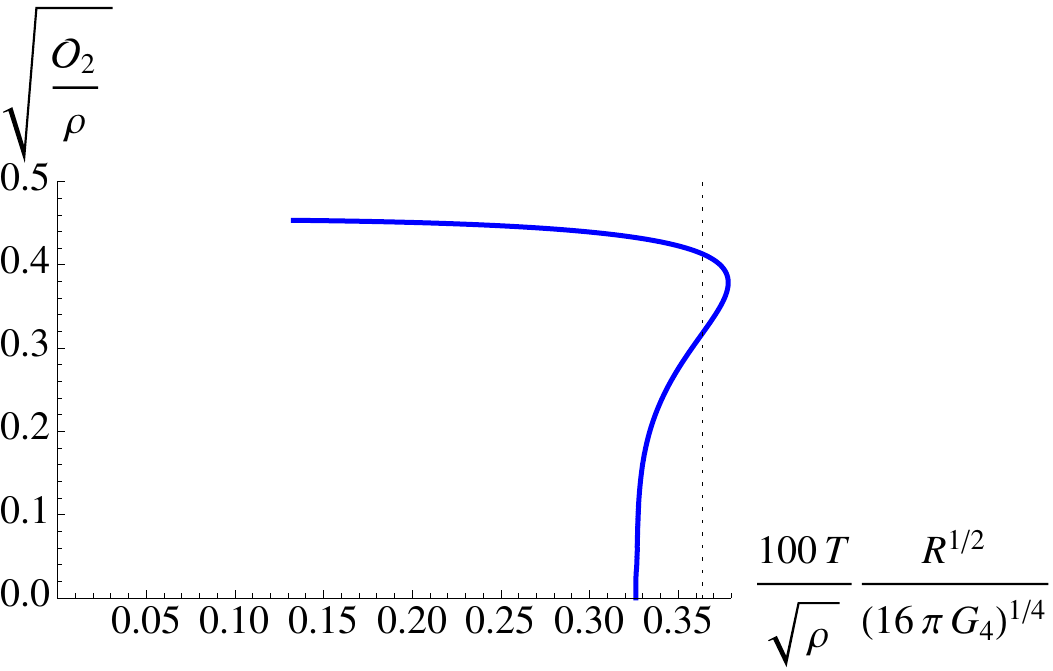}\label{fig:O2_vs_T}}
   \caption{\small Plots of operator versus temperature  for scalar charged black hole solutions with  either $\mathcal{O}_1$ or $\mathcal{O}_2$ non--zero. The vertical dotted line on the $\mathcal{O}_2$ plot denotes the transition temperature. See text.}  \label{fig:operator}
   \end{center}
\end{figure}

The solutions can only be exhibited numerically, and we display the resulting plots of  temperature versus operator vev for each case of  $\mathcal{O}_2$ and $\mathcal{O}_1$  in figure \ref{fig:operator}. Below a  critical temperature~$T_c$,  this type of solution is thermodynamically favored over the Reissner--Nordstr\"om case, and represents the superconducting phase, with non--zero condensate. These thermodynamics will be reviewed in the next section.
%
\subsection{Zero Temperature Phase}
%
The zero temperature solution is an RG flow between two AdS spaces\cite{Bobev:2011rv}.  In the IR, the fields have the behaviour:
\begin{equation}
\lambda(\tilde{z}) = \ln \left( 2 + \sqrt{5} \right) + \lambda_1 \tilde{z}^{-\alpha} + \dots \ , \quad \psi(\tilde{z}) = \psi_1 \tilde{z}^{- \beta} + \dots \ , \quad f(\tilde{z}) = \frac{7}{3} + \dots \ , \quad \chi(\tilde{z}) = \chi_0 + \dots \ ,
\end{equation}
where
\begin{equation}
\alpha = \sqrt{ \frac{303}{28} } - \frac{3}{2} \ , \quad \beta = \sqrt{ \frac{247}{28}} - \frac{1}{2} \ .
\end{equation}
%

\subsection{Thermodynamics}
\begin{figure}[h]
\begin{center}
   \subfigure[$\mathcal{O}_1$]{\includegraphics[width=3.0in]{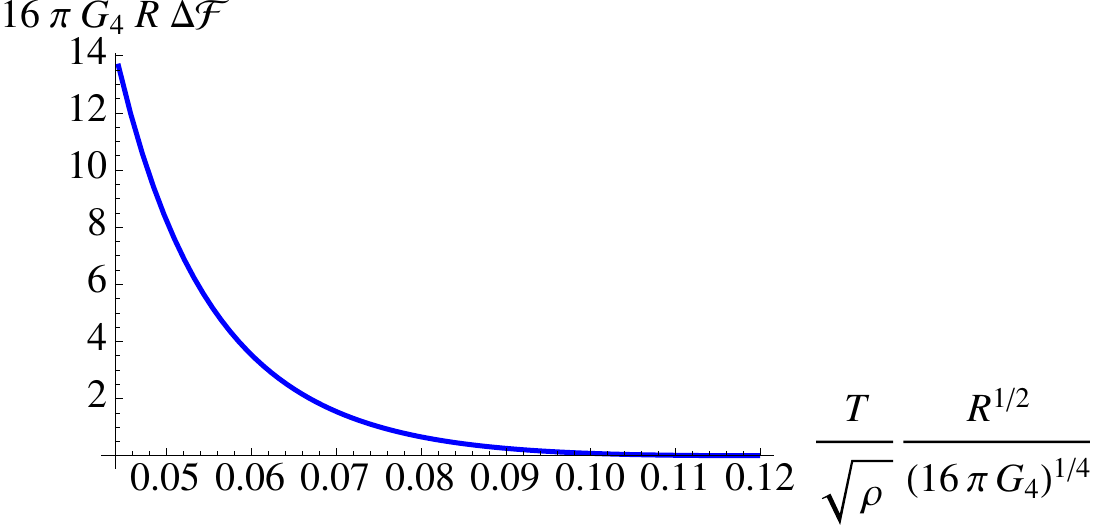}\label{fig:FreeEnergyO1}} \hspace{0.5cm}
\subfigure[$\mathcal{O}_2$]{\includegraphics[width=3.0in]{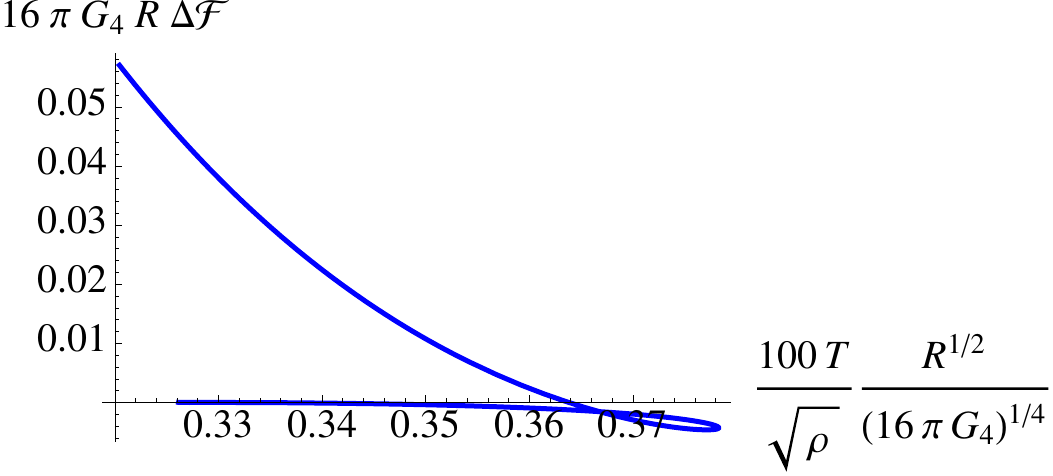}\label{fig:FreeEnergyO2}}
   \caption{\small Free energy density difference.  When $\Delta \mathcal{F} > 0$, the superconductor is thermodynamically favored.  This occurs at $T_c \approx 0.1199 \frac{\sqrt{\rho} (16 \pi G_4 )^{1/4}}{R^{1/2}}$ for the $\mathcal{O}_1$ case and $100T_c \approx 0.3638  \frac{\sqrt{\rho} (16 \pi G_4 )^{1/4}}{R^{1/2}}$ for the $\mathcal{O}_2$ case.}  \label{fig:FreeEnergy}
   \end{center}
\end{figure}

%
The on--shell regularized action is given by:
\begin{eqnarray}
I &=& I_{EH} + I_{\partial} + I_{CT} \ , \qquad {\rm with}\nonumber \\
I_{EH} &=& - \frac{1}{16 \pi G_4} \int d^4 x  \sqrt{G} \left( \mathcal{P} + \frac{1}{2} F_{\tilde{t} z}^2 |G^{zz}G^{\tilde{t}\tilde{t}}| \right) = \frac{V\beta R^2 e^{\chi_0/2}}{16 \pi G_4 } \int_\epsilon^{z_H} dz \left( - 2 \partial_z \left( \frac{f(z) e^{-\chi/2}}{z^3}\right) \right) \ , \nonumber \\
I_{\partial} &=& - \frac{1}{8 \pi G_4} \int d^3 x \sqrt{h} \mathcal{K} \ , \nonumber \\
I_{CT} &=& \frac{1}{8 \pi G_4} \int d^3 x \sqrt{h} \frac{2}{R}  - \frac{1}{16 \pi G_4} \int d^3 x \sqrt{h} \frac{2}{R} \lambda(\epsilon)^2 \ .
\end{eqnarray}
where we have used the equations of motion to simplify the on--shell action.  The quantity $V$  is the  volume of the ${\mathbb R}^2$ upon which the field theory resides. Putting everything in we get~\cite{Bobev:2011rv}:
\begin{equation}
I =  \left(\frac{\beta}{R} \frac{V}{16 \pi G_4} \left( f_3 - 4 \lambda_1 \lambda_2 \right) \right) = \beta V\mathcal{G} \ ,
\end{equation}
where $\mathcal{G}$ is the Gibbs energy density.  So we define the free energy density as:
\begin{equation}
\mathcal{F} =\mathcal{G} +  \rho \mu = \frac{1}{16 \pi G_4 R} \left( f_3 - e^{\chi_0} \psi_0 \psi_1 \right) \ .
\end{equation}
The Reissner--Nordstr\"om free energy density is given by:
\begin{equation}
\mathcal{F} _{RN} = \frac{1}{16 \pi G_4 R} \left( - \frac{1}{\tilde{z}_0^3} \left(1 + Q^2 \right) + 4 \frac{Q^2}{\tilde{z}_0^3} \right) \ ,
\end{equation}
where $(z_0, Q)$ are found by solving:
\begin{equation}
\frac{1}{4 \pi \tilde{z}_0} \left( 3 - Q^2 \right) = R T \ , \quad \frac{2 Q}{\tilde{z}_0^2} = - \psi_1 e^{\chi_0/2} \ .
\end{equation}
We define the difference of the free energy densities:
\begin{equation}
\Delta \mathcal{F} = \mathcal{F}_{RN} - \mathcal{F} \ .
\end{equation}
We show the free energy density difference as  function of temperature in figure \ref{fig:FreeEnergy}, for each case of $\mathcal{O}_1$ and $\mathcal{O}_2$. When $\Delta\mathcal{F}>0$, there is a transition from Reissner--Nordstr\"om to the black hole with scalar profile, representing the superconducting phase.  This defines the phase transition temperature, $T_c$.

Note that in the case of $\mathcal{O}_2$, it is the upper branch (the choice with higher vev for $\mathcal{O}_2$) that is favoured.  Here, in contrast to the $\mathcal{O}_1$ case where the vev rises continuously from zero at $T_c$, the $\mathcal{O}_2$ operator has a jump in the vev at $T_c$. The physics of these cases is described more in ref.\cite{Bobev:2011rv}.

It is worth noting here that while the  phase structure in the $\mathcal{O}_2$ case seems strikingly different from that of $\mathcal{O}_1$, we are aware of another model in the literature that shows how the two can be connected, although using an apparently different mechanism. In refs.~\cite{Basu:2008st,Herzog:2008he}, the introduction of a background current  in the $\mathbb{R}^2$ can continuously deform the solution space of the  $\mathcal{O}_1$ into that of the  $\mathcal{O}_2$ case. Here, instead of a current, the unusual behaviour of the metric function $f(z)$ is responsible for the multivaluedness of available scalar black holes for some ranges of temperatures. In either way of thinking about it, there is a new length scale in the theory that manifests itself as a finite value in the jump of  the free energy and of $\mathcal{O}_2$ as the transition temperature is crossed. As we will see, the entanglement entropy will be able to detect this new length scale.

%
\section{Entanglement Entropy}
%
With a complete  supergravity  solution in hand, we are ready to study the entanglement entropy at each temperature and study the physics.. We use a strip geometry, as outlined in the introduction.  We choose the following  embedding:
\begin{equation}
\xi^1 = x \ , \quad \xi^2 = y \ , \quad z = z(x) \ .
\end{equation}
The resulting entanglement entropy is given by:
\begin{equation}
4 G_4 S = L \int_{-\ell/2}^{\ell/2} d x \frac{R^2}{z^2} \left(1 + \frac{z'(x)^2}{f(z)} \right)^{1/2} \ .
\end{equation}
The extremization problem has a  constant of the motion given by:
\begin{equation}
\frac{1}{z_\ast^2} = \frac{1}{z^2} \frac{1}{\sqrt{1 + \frac{z'(x)^2}{f(z)}}} \ ,
\end{equation}
where $z_\ast$ is the location in $z$ of the bottom of the extremal surface.  This allows us to write the entanglement entropy as:
\begin{equation}
4 G_4 S = 2 L R^2 \int_\epsilon^{z_\ast} d z \frac{z_\ast^2}{z^2} \frac{1}{\sqrt{ \left( z_\ast^4 - z^4 \right) f(z)} } = 2 L R^2 \left( s + \frac{1} {\epsilon} \right) \ ,
\end{equation}
where $s$ has dimensions of inverse length with no divergences.  The length $\ell$ is given in terms of $z_\ast$:
\begin{equation}
\frac{\ell}{2} = \int^{z_\ast}_{\epsilon} d z \frac{z^2}{\sqrt{\left( z_\ast^4 - z^4 \right) f(z)}} \ .
\end{equation}
Under this scaling of equation \reef{eqt:scaling3}, $\ell$ and $s$ scale as:
\begin{equation}
\ell \to \gamma_3^{-1} \ell \ , \quad s \to \gamma_3 s \ ,
\end{equation}
so we will focus on the following dimensionless quantities:
\begin{equation}
\sqrt{\rho} \ \ell \ , \quad \frac{s}{\sqrt{\rho}} \ .
\end{equation}
We are now ready to explore the results.
%
\subsection{$\mathcal{O}_1$ Superconductor}\label{sec:EEsupercon_1}
\begin{figure}[h]
\begin{center}
\subfigure[$\frac{R^{1/2}}{\left( 16 \pi G_4 \right)^{1/2}} \frac{T}{\sqrt{\rho}}  = 0.07019$]{\includegraphics[width=3.0in]{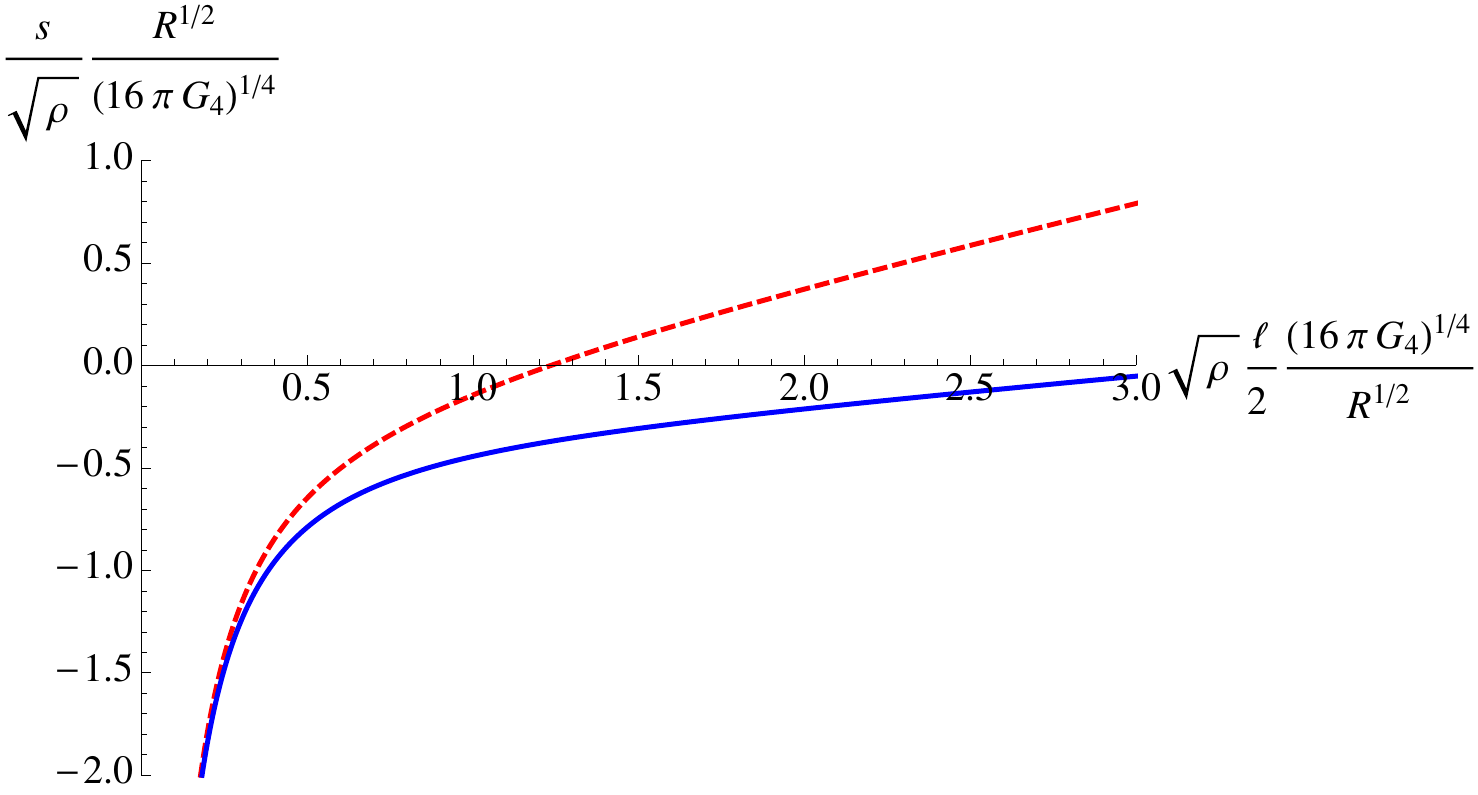}\label{fig:EE_O1_i=200}} \hspace{0.8cm}
\subfigure[$\frac{R^{1/2}}{\left( 16 \pi G_4 \right)^{1/2}} \frac{T}{\sqrt{\rho}}  = 0$]{\includegraphics[width=3.0in]{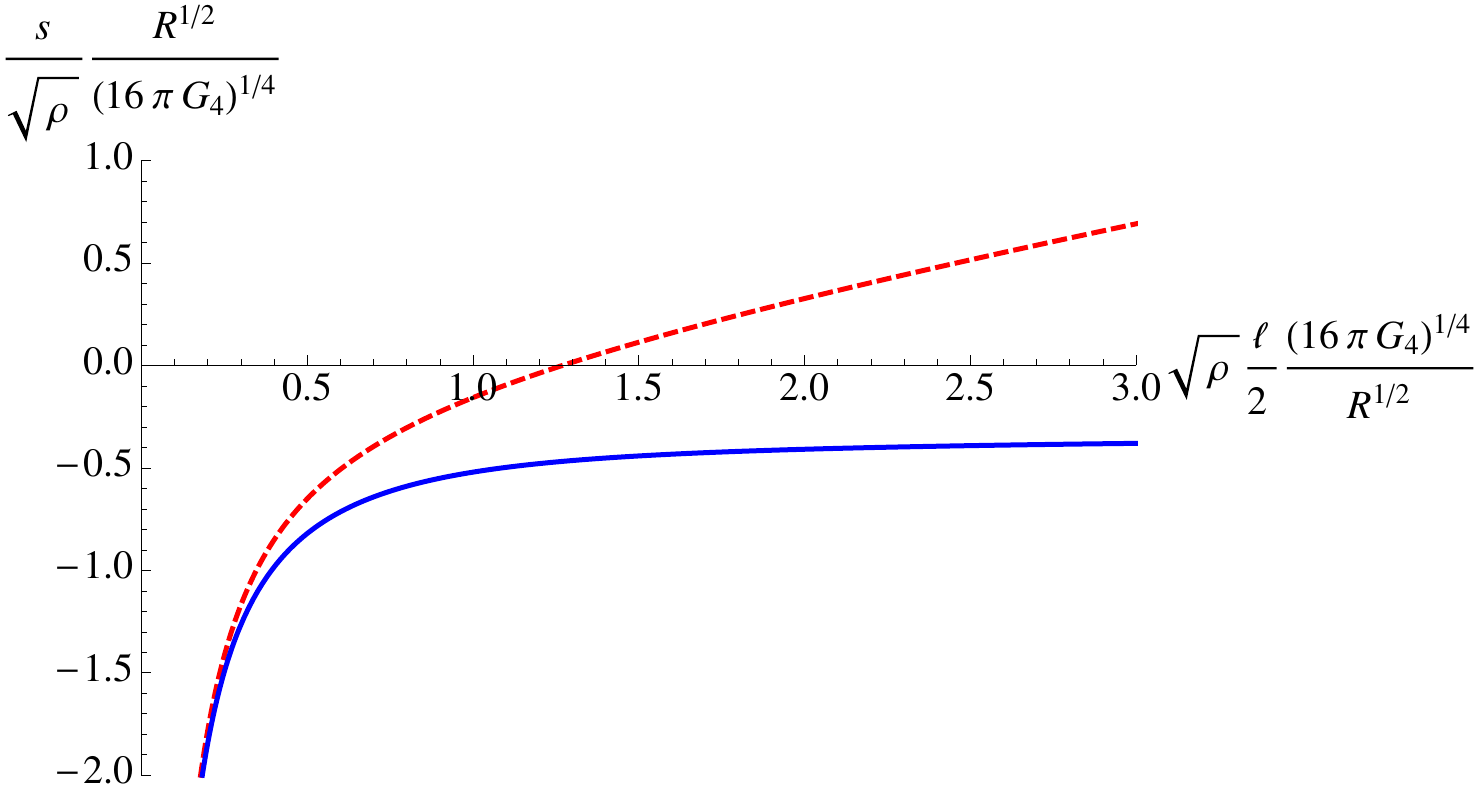}\label{fig:EE_O1_T=0}}
   \caption{\small Entanglement entropy {\it vs.} strip width $\ell$ for the $\mathcal{O}_1$ case. The solid blue curve is the superconductor solution, and the red dashed curve is the Reissner--Nordstr\"om solution.}  \label{fig:EE_O1}
   \end{center}
\end{figure}

%
We show in figure \ref{fig:EE_O1} the results for $s$ obtained by fixing the temperature and varying $\ell$,  the width of the strip. Larger $\ell$ probes more deeply into the infra--red. We show cases with temperature below the transition temperature $T_c$.   (At the transition temperature, the curves for the Reissner--Nortstr\"om case and the scalar charged black hole (representing the superconducting phase) are identical.)  In all cases, the curves go  linearly with $\ell$ for large $\ell$ as is expected from the area law.  As the temperature is lowered, the slope of the curve for large $\ell$ (still linear) is smaller for the superconducting background, and continues to flatten out as we approach zero temperature.  This is expected since for $T=0$, the background is an RG flow from one AdS  vacuum to another, and this flattening out of the entanglement entropy at some finite value was observed in our studies of entanglement entropy along RG flow presented in  ref.~\cite{Albash:2011nq}.  

The fact that for a particular  $\ell$, the superconducting solution exhibits a lower entropy than the Reissner--Nordstr\"om solution fits with the expectation that degrees of freedom have condensed and so there should be fewer of them.

\begin{figure}[h]
\begin{center}
  \includegraphics[width=4.0in]{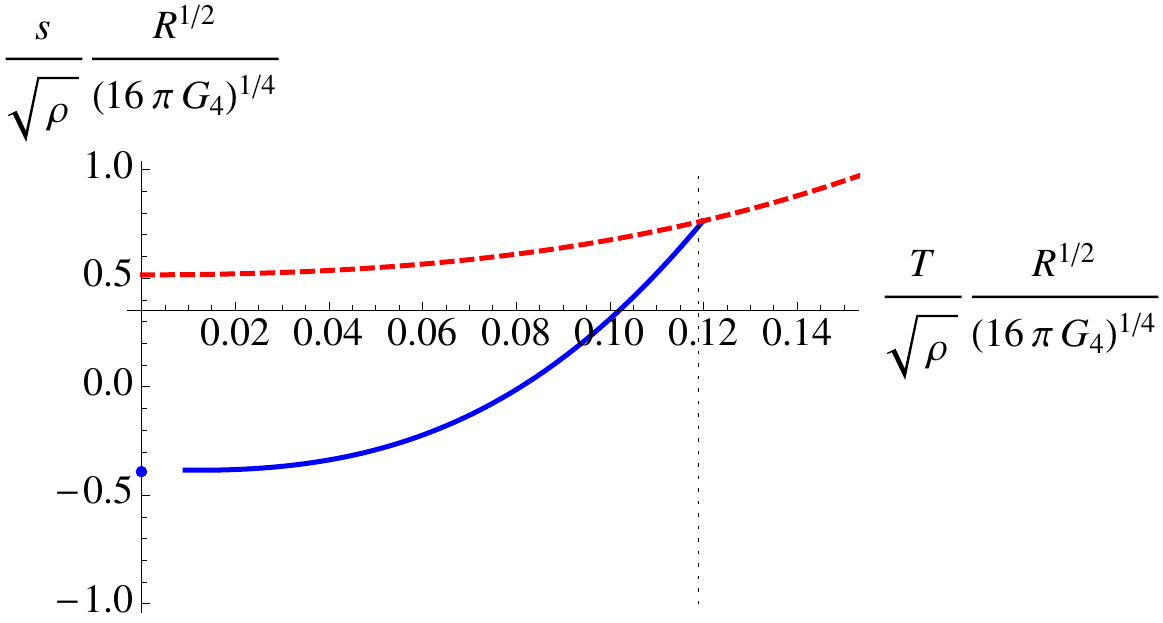}
   \caption{\small The entanglement entropy   in the $\mathcal{O}_1$ case,  as a function of  temperature, for fixed $\ell$. (We choose $\sqrt{\rho}(16 \pi G_4)^{1/4}R^{-1/2} \ell/2 = 2.5$) The solid blue curve is from  the superconductor solutions, while the  red dashed curve is from the Reissner--Nordstr\"om solutions. Trace the physical curve by always choosing the lowest entropy at a given $T$.  There is a  discontinuity in the slope of the decreasing entanglement entropy at the transition temperature $T_c$, indicated by the vertical dotted line. (While we do not plot all the superconductor points, due to lack of numerical control at low temperature, we display the zero temperature solution, since the solution is known exactly there.) }  \label{fig:EE_O1_fix_l}
   \end{center}
\end{figure}

 It is instructive to  slice the data differently, fixing a strip width $\ell$ and studying how the entropy of the system changes with temperature. We present this in figure~\ref{fig:EE_O1_fix_l}.  In reading the figure,  determine the physical curve by always choosing the point of lowest entropy at a given $T$. A  discontinuity in the slope of the decreasing entanglement entropy can be observed at the transition temperature (indicated by the vertical dotted line), showing its utility as an independent probe of the phase structure of the superconductor. 
 
 The slope may be thought of as a sort of response function  characterizing the system, roughly analogous to a specific heat. It is natural for it to be positive, since increasing temperature should indeed promote entanglement entropy. A discontinuous change in the slope at the transition temperature $T_c$ signifies a significant reorganization of the degrees of freedom of the system.  Since there is a condensate generated, it is also to be expected that there is a reduction in the number of degrees of freedom as well, although precisely at $T=T_c$, the condensate value only just begins to rise from zero. We will see something more dramatic in the $\mathcal{O}_2$ case, by way of contrast.

%

%
\subsection{$\mathcal{O}_2$ Superconductor}\label{sec:EEsupercon_2}
\begin{figure}[h]
\begin{center}
\subfigure[$\frac{R^{1/2}}{\left( 16 \pi G_4 \right)^{1/2}} \frac{100T}{\sqrt{\rho}}  = 0.320$]{\includegraphics[width=3.0in]{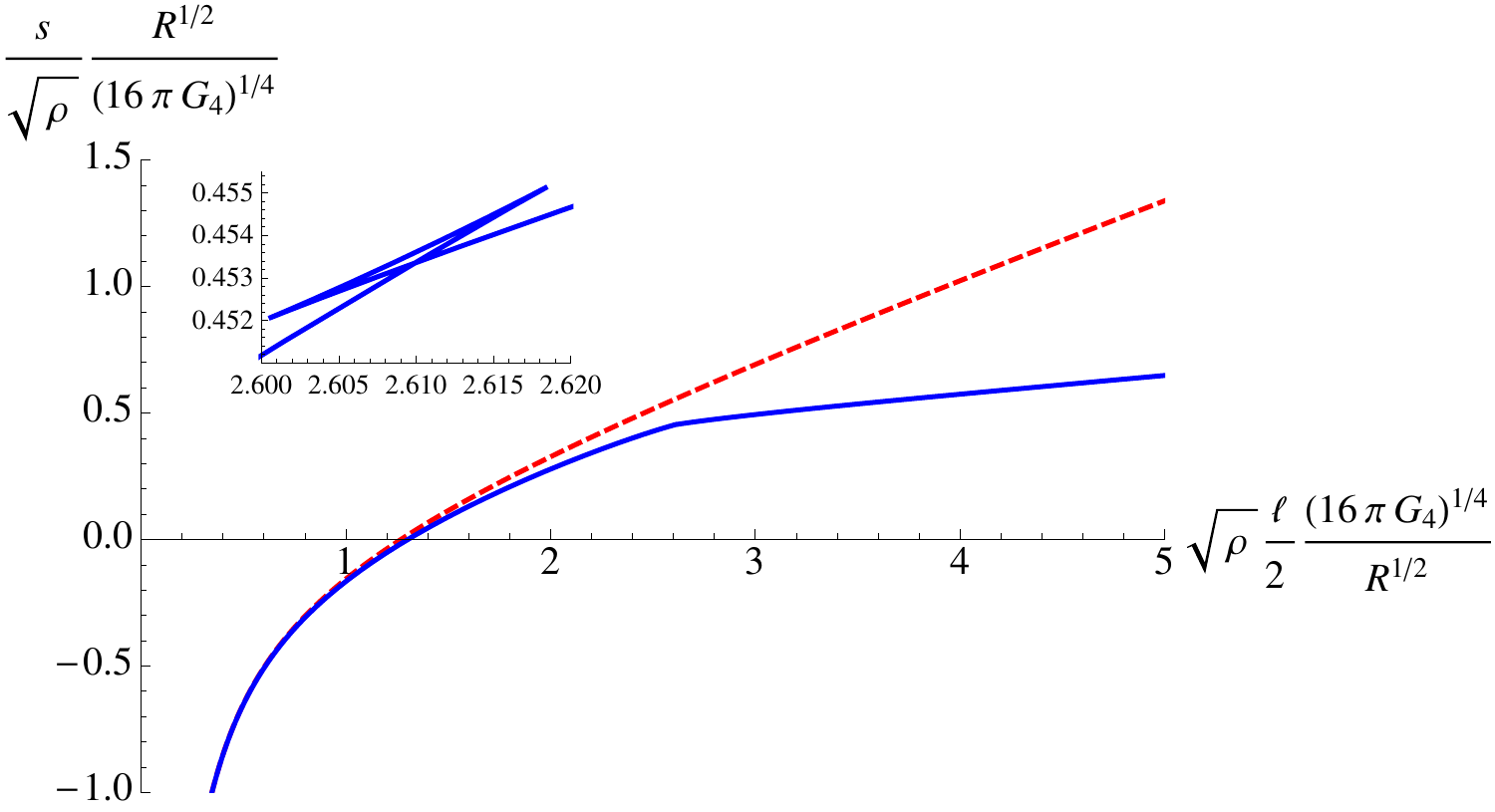}\label{fig:EE_O2_i=300}} \hspace{0.5cm}
\subfigure[$\frac{R^{1/2}}{\left( 16 \pi G_4 \right)^{1/2}} \frac{100T}{\sqrt{\rho}}  = 0$]{\includegraphics[width=3.0in]{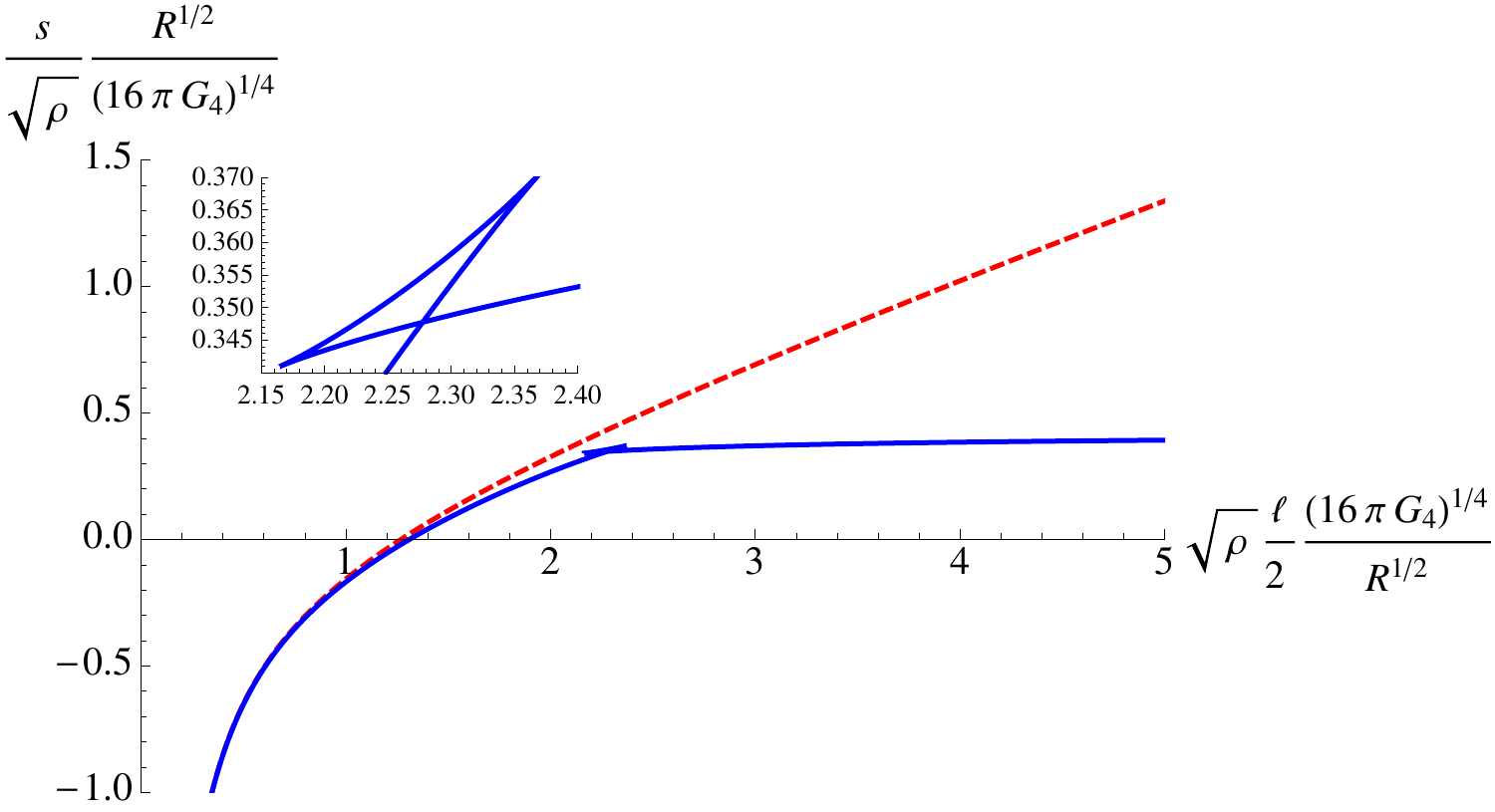}\label{fig:EE_O2_T=0}}
   \caption{\small The entanglement entropy for the $\mathcal{O}_2$ case.  The solid blue curve is the superconductor solution, red dashed curve is the Reissner--Nordstr\"om solution. There is a multivaluedness at finite $\ell$ in the shape of a swallowtail curve. At lower $T$, it occurs at lower $\ell$, and at $T=0$ the swallowtail region persists at finite $\ell$.}  \label{fig:EE_O2}
   \end{center}
\end{figure}

%
We show in figure \ref{fig:EE_O2} the results for $s$ obtained by fixing the temperature and varying $\ell$, the width of the strip. As before, larger $\ell$ probes more deeply into the infra--red. We again show cases  with temperature below the transition temperature $T_c$, although in this case, at $T_c$, the Reissner--Nordstr\"om  curve lies above that of the superconductor curve. This will mean a discontinuous jump in the {\it value} of the entropy, as we will see below, in contrast to the $\mathcal{O}_1$ case.   For all cases,    the typical behavior of the curve initially resembles that of the $\mathcal{O}_1$ case, in that the curve of the superconducting solution has a lower slope.  However, for $T<T_c$, we find multiple solutions for a given range of strip widths $\ell$, which form a swallowtail shape in our curves.  This means there is a kink in the physical curve for the entanglement entropy, since we must choose the lowest value.  For future reference, we will refer to the two parts of such kinked $(s,\ell)$  curves as the ``small $\ell$" branch and the ``large $\ell$" branch, respectively, going from small to large $\ell$.

The kink moves to lower strip widths $\ell$ as the temperature is decreased (for fixed charge density), and persists at zero temperature.  Another interesting point is that the leveling off of the curve at zero temperature occurs at a positive finite value.  This is interesting since in ref.~\cite{Albash:2011nq}, only negative finite values were observed. Furthermore,  the $\mathcal{O}_1$ case of the previous subsection also exhibits a negative finite value.  In ref.~\cite{Albash:2011nq}, we predicted from our sharp domain wall analysis that a positive finite value would develop if the domain wall was sufficiently sharp and near the UV. We will explore this in  subsection~\ref{sec:domain}.

\begin{figure}[ht] 
   \centering
   \includegraphics[width=3.5in]{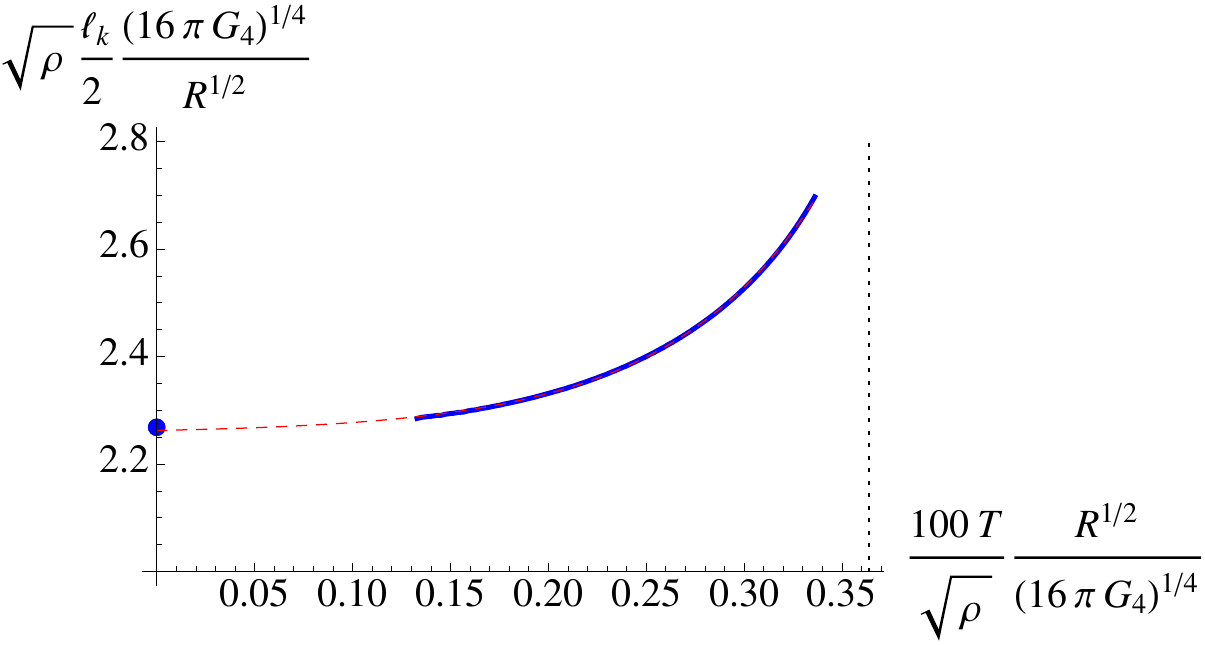} 
   \caption{\small The position of the kink in the entanglement entropy as a function of temperature.  The red dashed line is the best fit curve given by $a e^{b x} + c$,  with $a=0.00553$, $b=12.9848$, and $c=2.25678$. The curve exists between an $\ell_{\rm min}\simeq 2.27$ and an $\ell_{\rm max}\simeq 2.70$, discussed in the text. The vertical line indicates the position of $T_c$, above which there is no kink.}
   \label{fig:Lambda}
\end{figure}

It is also interesting to  track, as a function of temperature,   the strip width value, $\ell_k$, at which  the kink appears in the entanglement entropy.   We do this in figure \ref{fig:Lambda}.  (Note also that, for fixed temperature, $\ell_k \propto \rho^{-1/2}$.) From this curve we can read off an interesting piece of information. If we pick a specific strip width value $\ell$ between an $\ell_{\rm min}\simeq 2.27$ and an $\ell_{\rm max}\simeq 2.70$  we can read off  a specific temperature at which the entanglement entropy is crossing over from the ``small $\ell$" branch of an $(s,\ell)$ curve  to the ``large $\ell$" branch of the $(s,\ell)$ curve, the branches being separated by a kink.  The significance of this temperature will be apparent shortly.

A swallowtail multivaluedness in the entanglement entropy, showing multiple extremal surface solutions at a given $\ell$, was first observed and characterized in our studies, presented in ref.~\cite{Albash:2010mv}, of the evolution of the entanglement entropy after a quenching process\footnote{Refs.\cite{Balasubramanian:2011ur,Myers:2012ed} have also since observed this phenomenon.}.   We will  discuss the  origins of the multivaluedness of the present case in the next subsection.

Physically, the appearance of a kink in the entanglement entropy as we go to larger $\ell$ can be attributed to sensitivity to a new scale  in the theory, and the entanglement entropy is a good probe of its presence. In contrast to the $\mathcal{O}_1$ case, the transition at $T_c$ was associated with a finite jump in the free energy $\mathcal{F}$, and  also a jump in the the vev  of the operator $\mathcal{O}_2$. This sets and additional scale in the $\mathcal{O}_2$ theory that distinguishes it from the $\mathcal{O}_1$ case. (See also our discussion, near the end of section 2.4, of this  scale and how it can arise from the addition of  a background current  in the probe limit.) The basic scale, which we can denote $\tilde\xi$, is set by the inverse of   the discontinuity of the vev of $\mathcal{O}_2$ at $T=T_c$. For subsequent temperature $T<T_c$ an effective scale $\xi<{\tilde\xi}$ follows from this, by RG flow.  For very small strip size $\ell$, the entanglement entropy will not be sensitive to $\xi$, but when $\ell$ becomes comparable to $\xi$, our results suggest that correlations between quanta on these scales effectively reduce the  number of effective degrees of freedom,  reducing the contributions to the entanglement entropy, resulting in a kink to change of the slope of the $(s,\ell)$ curve for larger~$\ell$. The details of how this works from the field theory perspective, and the kink's fate away from the large $N$ limit, would be interesting to explore further, in future work.

As was done for the case of $\mathcal{O}_1$, it is instructive to study the entropy for fixed $\ell$ as the temperature varies.  There are in fact three distinct situations, giving rise to three different types of curve. The distinguishing  issue is whether the choice of fixed $\ell$ can ever become a kink value, $\ell_k$,  at some temperature. As we saw, figure~\ref{fig:Lambda} has the answer to this. 

\begin{figure}[h]
\begin{center}
\includegraphics[width=4.0in]{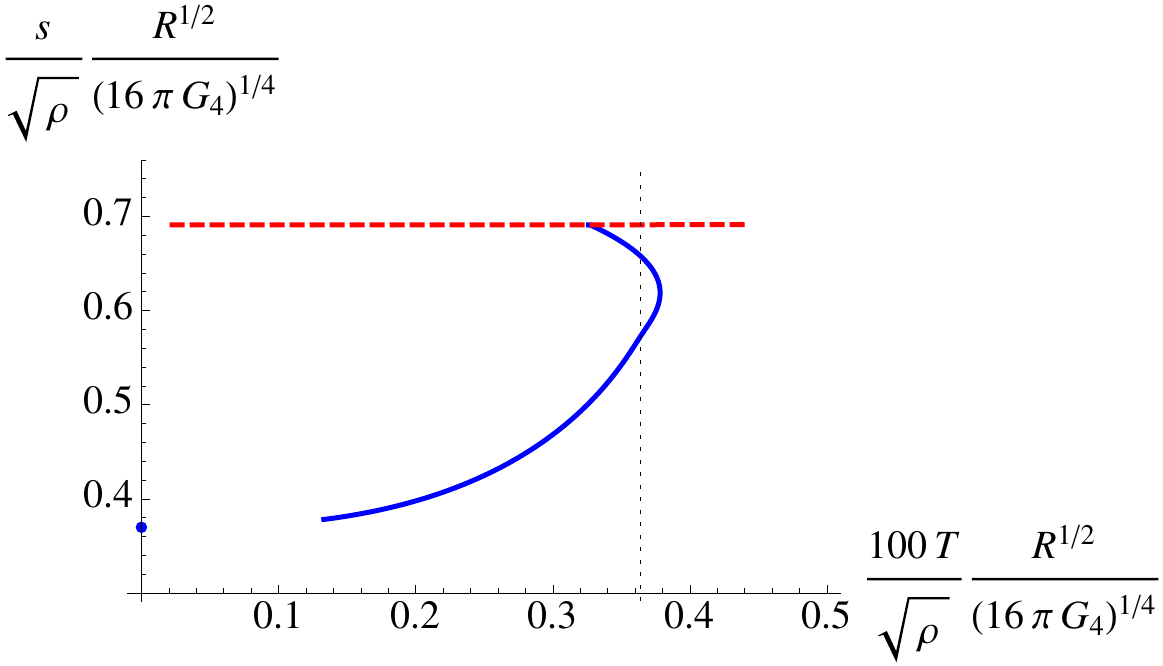}
   \caption{\small The entanglement entropy  in the $\mathcal{O}_2$ case,  as a function of  temperature, for fixed $\ell$. (We choose $\sqrt{\rho}(16 \pi G_4)^{1/4}R^{-1/2} \ell/2=3$ here.) The solid blue curve is from  the superconductor solutions, red dashed curve  (decreasing in slope, but only slowly on this scale) is from  the Reissner--Nordstr\"om solutions.   Trace the physical curve by choosing the red curve for $T>T_c$, indicated by the vertical dotted line, and for $T<T_c$, always choosing the lowest entropy. There is therefore a discontinuous jump in $s$ and its slope  at  $T_c$. (While we do not plot all the superconductor points, due to lack of numerical control at low temperature, we display the zero temperature solution, since the solution is known exactly there.) }  \label{fig:EE_O2_fix_la}
   \end{center}
\end{figure}

The first case is to have  a fixed $\ell$ that is greater than $\ell_{\rm max}$. Then, at  successively lower temperatures than $T_c$, the entropy will always come from points on ``large $\ell$" branches of $(s,\ell)$ curves, since the kink moves to smaller $\ell$ as $T$ is reduced. We show the resulting $(s,T)$ type of curve in figure~\ref{fig:EE_O2_fix_la}. In this and the next two figures, the red curve is always favoured for  $T>T_c$, indicated by the vertical dotted line. Below $T_c$, one should determine the physical curve by always choosing the point of lowest entropy at a given $T$. Notice how the region of negative slope is nicely avoided. This will be the case in our subsequent curves as well.

\begin{figure}[h]
\begin{center}
\includegraphics[width=4.0in]{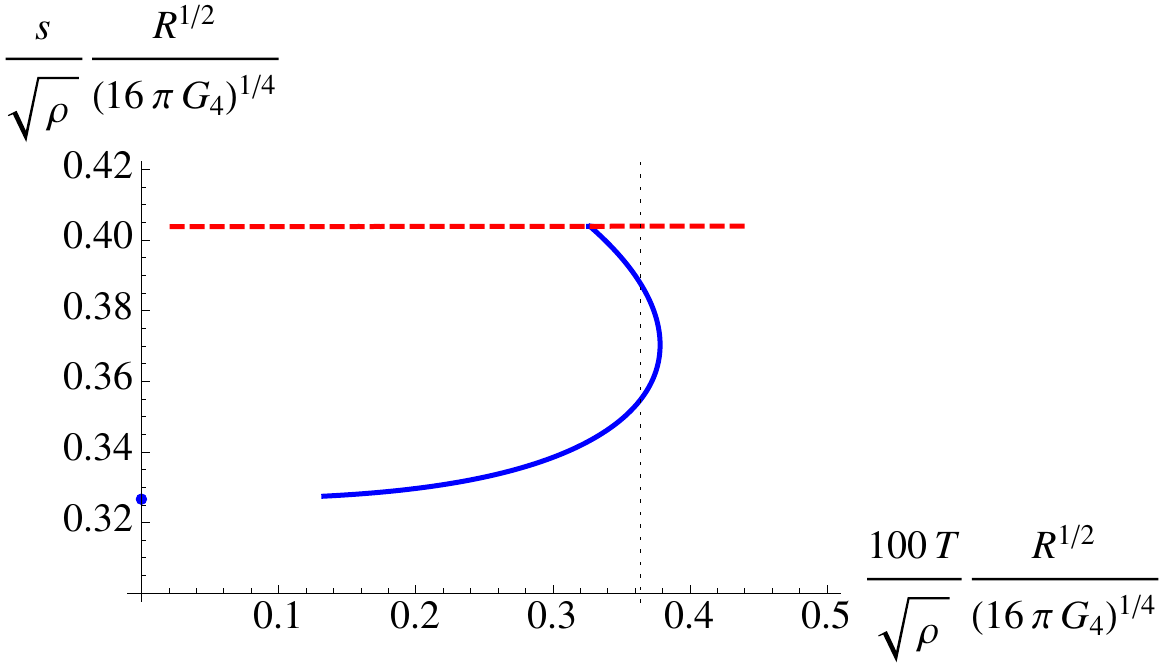}
   \caption{\small The entanglement entropy  in the $\mathcal{O}_2$ case,  as a function of  temperature, for fixed $\ell$. (We choose $\sqrt{\rho}(16 \pi G_4)^{1/4}R^{-1/2} \ell/2=2.2$ here.) The solid blue curve is from  the superconductor solutions, red dashed curve  (decreasing in slope, but only slowly on this scale) is from  the Reissner--Nordstr\"om solutions.  Trace the physical curve by choosing the red curve for $T>T_c$, indicated by the vertical dotted line, and for $T<T_c$, always choosing the lowest entropy. There is therefore a discontinuous jump in $s$ and its slope  at  $T_c$.  (While we do not plot all the superconductor points, due to lack of numerical control at low temperature, we display the zero temperature solution, since the solution is known exactly there.)}  \label{fig:EE_O2_fix_lb}
   \end{center}
\end{figure}

The second case is to have a fixed $\ell$ that is smaller than $\ell_{\rm min}$. There is no temperature at which this would become a kink value, since although the kink moves to smaller $\ell$ with smaller $T$, it stops at $\ell_{\rm min}$ at $T=0$. So the contributions to the entropy come entirely from points on ``small $\ell$" branches of $(s,\ell)$ curves. We show the resulting $(s,T)$ type of curve in figure~\ref{fig:EE_O2_fix_lb}.

\begin{figure}[h]
\begin{center}
  \includegraphics[width=4.0in]{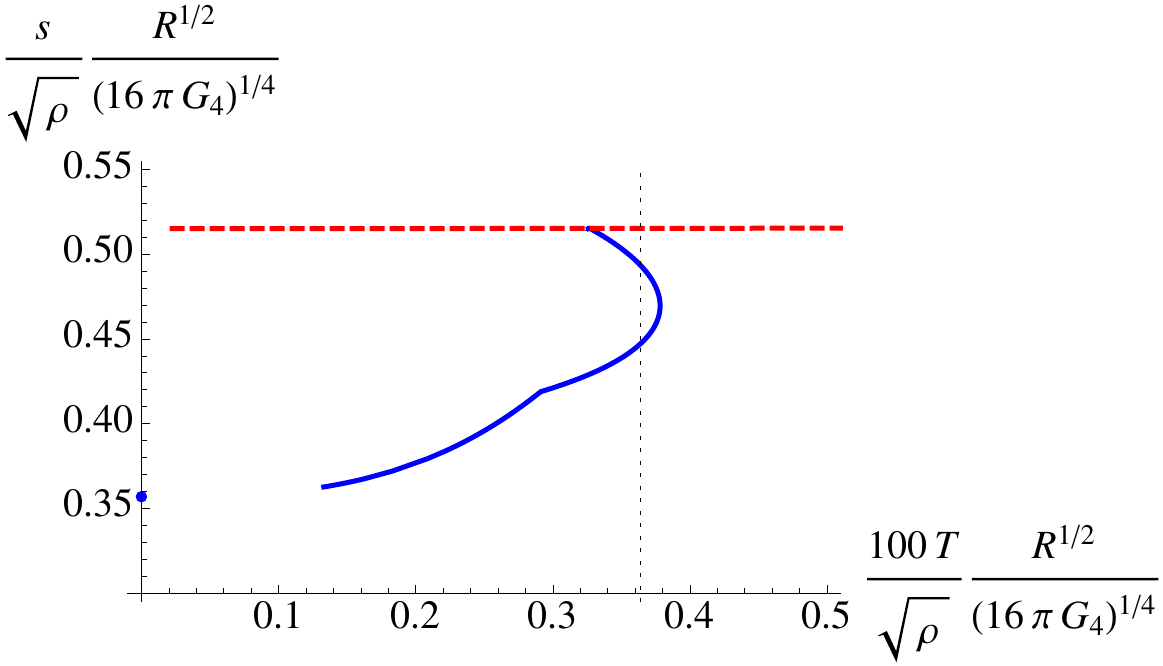}
   \caption{\small The entanglement entropy  in the $\mathcal{O}_2$ case,  as a function of  temperature, for fixed $\ell$. (We choose $\sqrt{\rho}(16 \pi G_4)^{1/4}R^{-1/2} \ell/2=2.5$ here.) The solid blue curve is from  the superconductor solutions, red dashed curve  (decreasing in slope, but only slowly on this scale) is from  the Reissner--Nordstr\"om solutions.    Trace the physical curve by choosing the red curve for $T>T_c$, indicated by the vertical dotted line, and for $T<T_c$, always choosing the lowest entropy. There is therefore a discontinuous jump in $s$ and its slope  at  $T_c$.  There is an additional jump in the slope at a  lower temperature. (While we do not plot all the superconductor points, due to lack of numerical control at low temperature, we display the zero temperature solution, since the solution is known exactly there.) }  \label{fig:EE_O2_fix_lc}
   \end{center}
\end{figure}

The final case is to have a fixed $\ell$ such that $\ell_{\rm min}\leq \ell\leq\ell_{\rm max}$. Then, as we reduce the temperature from $T_c$,  entropy contributions are from ``large $\ell$" branches until a temperature is reached such that our chosen $\ell$ is a kink value $\ell_k$. This  temperature can be read off from figure~\ref{fig:Lambda}.  For lower temperatures, the entropy will be from points on ``small $\ell$" branches. Consequently, the  superconductor phase of the $(s,T)$ curve in this case will be a combination of  two types of curve, connected by a new discontinuity  in the derivative where they join. We show the resulting $(s,T)$ type of curve in figure~\ref{fig:EE_O2_fix_lc}.

In all  the curves in figures~\ref{fig:EE_O2_fix_la}, \ref{fig:EE_O2_fix_lb} and figure~\ref{fig:EE_O2_fix_lc},    we  see that in  addition to  the slope having a discontinuity at the transition temperature $T_c$ (shown by the vertical dotted line),  the value of the entropy drops  discontinuously as well, as we earlier anticipated.  Then, when $\ell_{\rm min}\leq \ell\leq\ell_{\rm max}$, we have the additional feature of a discontinuity in the slope at some lower temperature, generated (as discussed above) by the sensitivity to the length scale $\xi$. Note that for fixed $\ell$ with a value close to the edges (but outside) of this interval, the resulting $(s,T)$ curves will appear to have a locally smoothed out discontinuity. The cases displayed here are far away enough from the interval that the smoothing is spread out.

\subsection{Multivaluedness of the Entanglement Entropy} 
\begin{figure}[h]
\begin{center}
   \subfigure[Finite Temperature]{\includegraphics[width=1.75in]{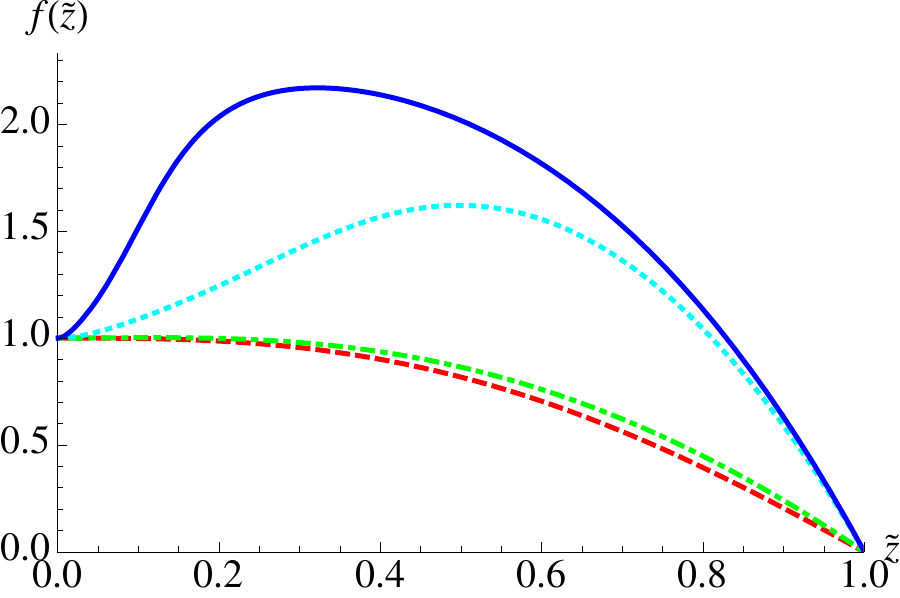}\label{fig:O1_f_T}} \hspace{0.5cm}
\subfigure[Zero Temperature]{\includegraphics[width=1.75in]{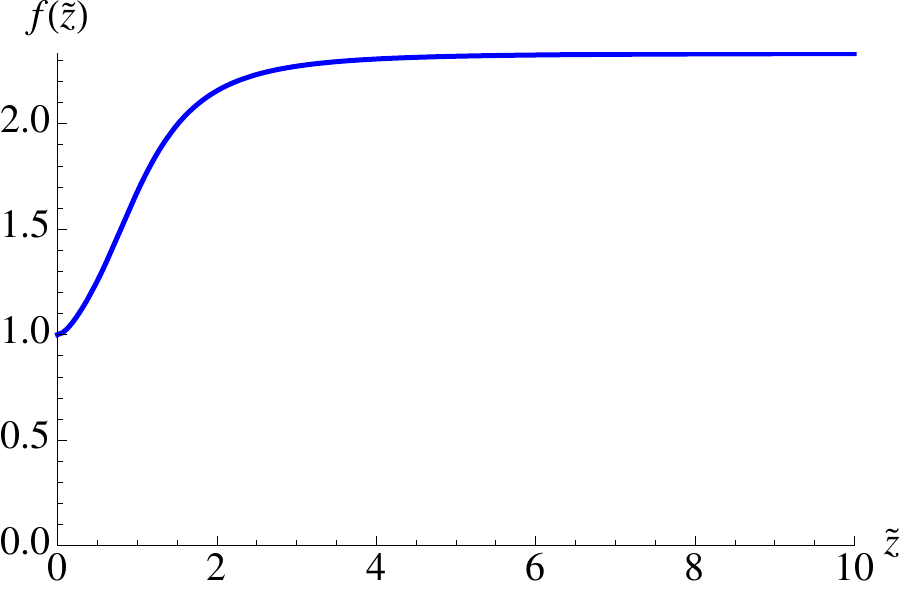}\label{fig:EE_f_T=0}}
   \caption{\small The metric function $f(z)$ for the $\mathcal{O}_1$ superconductor.  For (a), red dashed ($\frac{R^{1/2}}{\left( 16 \pi G_4 \right)^{1/2}} \frac{T}{\sqrt{\rho}}  =0.11993$), green dot--dashed ($ \frac{R^{1/2}}{\left( 16 \pi G_4 \right)^{1/2}} \frac{T}{\sqrt{\rho}}  =0.10309$), cyan dotted ($\frac{R^{1/2}}{\left( 16 \pi G_4 \right)^{1/2}} \frac{T}{\sqrt{\rho}} =0.02990$), solid blue ($\frac{R^{1/2}}{\left( 16 \pi G_4 \right)^{1/2}} \frac{T}{\sqrt{\rho}}  =0.00932$).}   \label{fig:O1_f}
   \end{center}
\end{figure}

\begin{figure}[h]
\begin{center}
   \subfigure[Finite Temperature]{\includegraphics[width=1.75in]{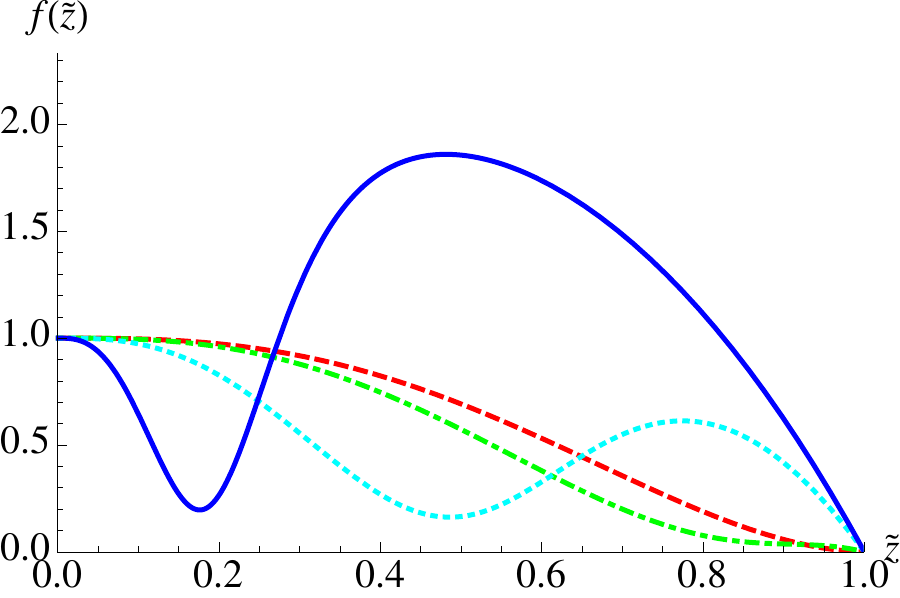}\label{fig:O2_f_T}} \hspace{0.5cm}
\subfigure[Zero Temperature]{\includegraphics[width=1.75in]{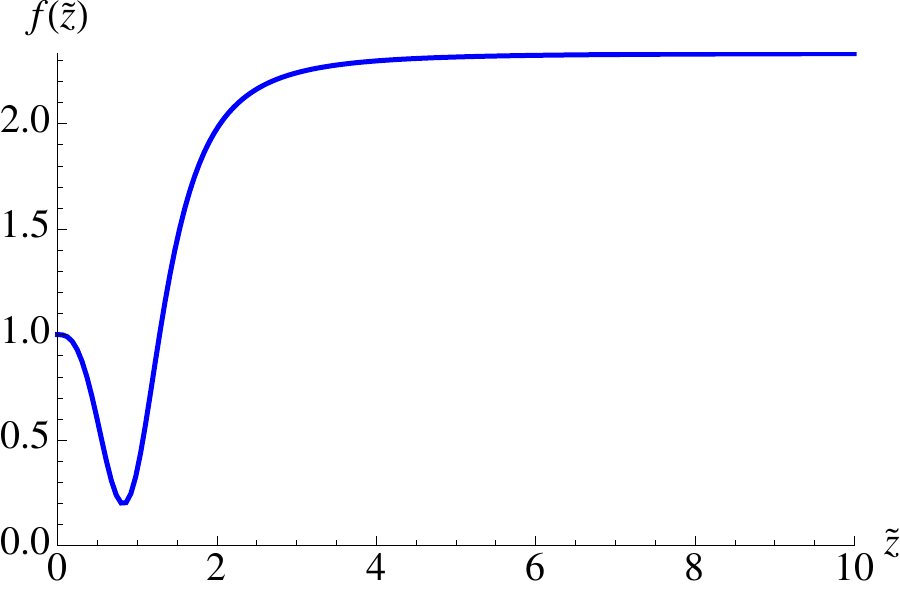}\label{fig:O2_f_T=0}}
   \caption{\small The metric function $f(z)$ for the $\mathcal{O}_2$ superconductor.  For (a), red dashed ($\frac{R^{1/2}}{\left( 16 \pi G_4 \right)^{1/2}} \frac{100T}{\sqrt{\rho}}  =0.326$), green dot--dashed ($ \frac{R^{1/2}}{\left( 16 \pi G_4 \right)^{1/2}} \frac{100T}{\sqrt{\rho}}  =0.369$), cyan dotted ($\frac{R^{1/2}}{\left( 16 \pi G_4 \right)^{1/2}} \frac{100T}{\sqrt{\rho}} =0.320$), solid blue ($\frac{R^{1/2}}{\left( 16 \pi G_4 \right)^{1/2}} \frac{100T}{\sqrt{\rho}}  =0.133$).}   \label{fig:O2_f}
   \end{center}
\end{figure}

As stated already, we've seen  a multivaluedness of the entanglement entropy before, in a study of its evolution after a quench~\cite{Albash:2010mv}. This resulted in a kink representing the change in the saturation rate of the entropy as it evolved.  To understand the reason for the appearance of the multivaluedness in this case, we study the behavior of the function $f(z)$, which we show examples of in figure \ref{fig:O2_f} (we show the behavior for the $\mathcal{O}_1$ function in figure \ref{fig:O1_f} for comparison purposes).  Slightly before the rightmost point in the vev curve of figure~\ref{fig:O2_vs_T}, $f(z)$ develops two new extrema, a minimum and a maximum.  The maximum grows higher in value and the minimum becomes sharper ({\it i.e.} greater second derivative) as the temperature decreases.  Both extrema move toward the AdS boundary at ${\hat z}=0$ (the UV) as the temperature decreases.  

\begin{figure}[h]
\begin{center}
\subfigure[$f(\tilde{z})$]{\includegraphics[width=2.0in]{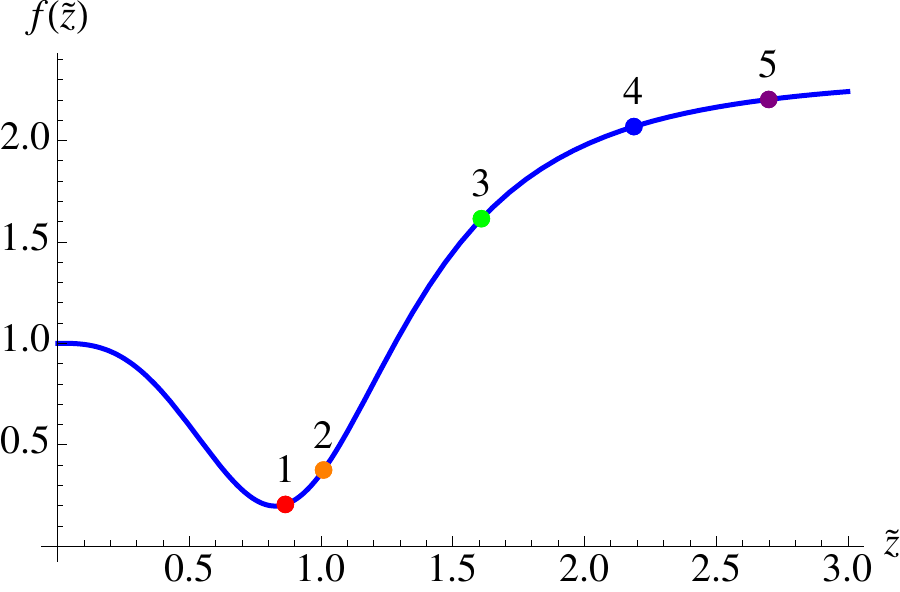}\label{fig:Numbered_f}}
 \hspace{1.5cm}
   \subfigure[Entanglement Entropy]{\includegraphics[width=2.6in]{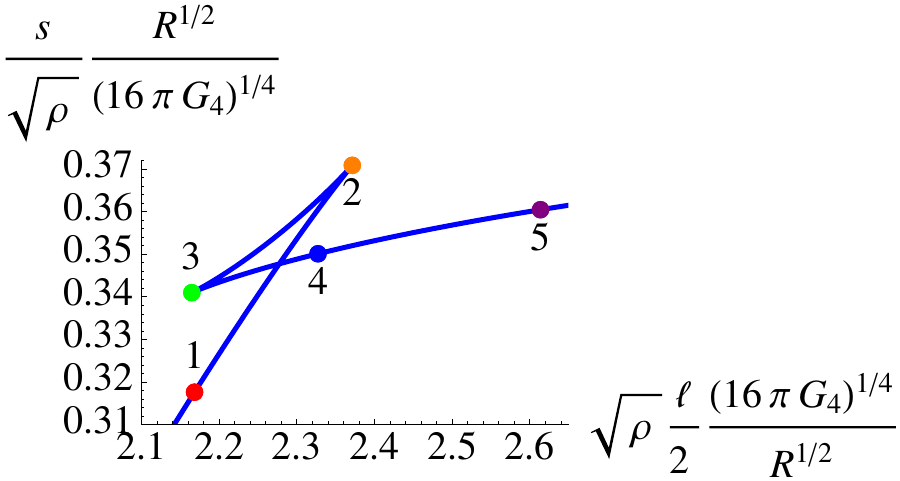}\label{fig:Numbered_S}} 
   \caption{\small A comparison of where the minimal surface corresponding to the entanglement entropy ends and the multivaluedness of the entanglement entropy.  The colored and numbered dots correspond to where the surface used to calculated the entanglement entropy ends in the bulk.  }   \label{fig:Numbered}
   \end{center}
\end{figure}

We show the behavior of the function $f(z)$ at zero temperature in figure \ref{fig:O2_f_T=0}.  As  for the finite (but low)  temperature case, the function has a minimum.  This minimum is now a finite distance from the UV boundary, and the maximum we saw at finite temperature has now smoothed out to a constant as we go toward  the IR.  
It is the non--monotonic behavior of $f(z)$ that generates the multivaluedness, as illustrated in figure \ref{fig:Numbered}.  When $z_\ast$, the location of the lowest point of the extremal surface, is in the neighborhood of  the minimum of $f(\tilde{z})$, the entanglement entropy becomes multivalued.  Only when relatively far from the minimum does the entanglement entropy become single--valued again.

\subsection{Domain Wall Behaviour}\label{sec:domain}

\begin{figure}[h]
\begin{center}
   \subfigure[]{\includegraphics[width=2.0in]{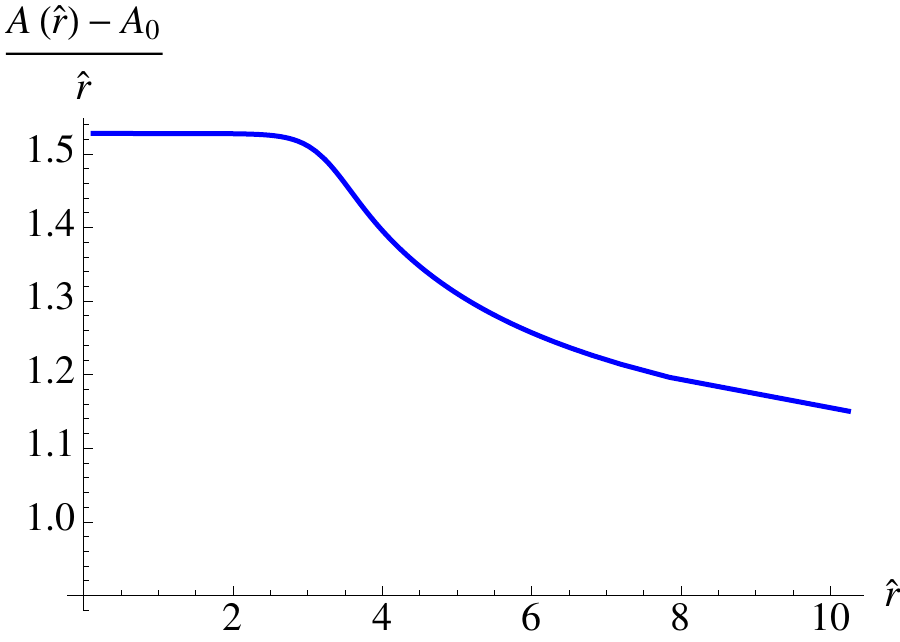}\label{fig:DomainWall1}} \hspace{2.5cm}
\subfigure[]{\includegraphics[width=2.0in]{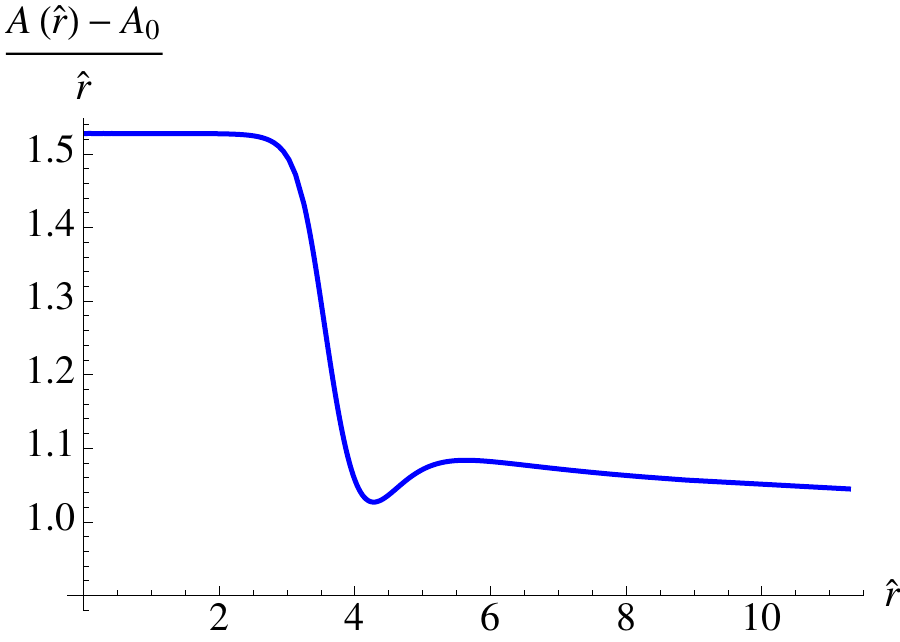}\label{fig:DomainWall2}}
   \caption{\small The domain wall for (a) the $\mathcal{O}_1$ case  and (b) the $\mathcal{O}_2$ case.  The UV is at  ${\hat r} \to \infty$, and the results have been shifted such that the IR results are at finite values of the radial coordinate.}   \label{fig:DomainWall}
   \end{center}
\end{figure}

The non--monotonic behavior of $f(\tilde{z})$ suggests that the behavior of the domain wall that interpolates between the two AdS vacua (at zero temperature) is also not monotonic.  In addition, we'd also like to compare the domain wall features of the cases $\mathcal{O}_1$ and $\mathcal{O}_2$, to see if we can understand the differences between the  large $\ell$ saturation values of the entanglement entropy observed in sections~\ref{sec:EEsupercon_1} and~\ref{sec:EEsupercon_2} (negative  {\it versus} positive) in the terms discussed in our RG flow studies of ref.~\cite{Albash:2011nq}. To study this, we first write the metric in the following form:
\begin{equation}
ds^2 = - e^{2 A(r)} e^{-\chi(z) + \chi(0)}dt^2 + \frac{R^2}{z(r)^2} d \vec{x}^2 + dr^2\ .
\end{equation}
The function $A(r)$ encodes the domain wall, and the coordinate $r$ can be determined via:
\begin{equation}
r(z) - r_0 = - \int_{\infty}^z \frac{dz' }{z' f(z')} \ .
\end{equation}
Note that technically $r_0$ is  actually at $-\infty$ since that corresponds to the IR of the theory. 
 So to circumvent this issue, we can define a variable $\hat{r}$ such that:
\begin{equation}
{\hat r}(z) = -  \int_{\tilde{z}_{max}}^{\tilde{z}} \frac{dz' }{z' f(z')} \ ,
\end{equation}
and simply shift our result for $A(\hat{r})$ by a constant $A_0$ such that the ratio $(A(\hat{r})-A_0)/{\hat r}$ does not diverge at ${\hat r}= 0$.  We show the behavior of $A({\hat r})$ in figure \ref{fig:DomainWall}.  Indeed, as expected, it is non--monotonic for the $\mathcal{O}_2$ case, and indeed the domain wall is much sharper for this case than it is for the $\mathcal{O}_1$ case, confirming our observations made in section~\ref{sec:EEsupercon_2} and ref.~\cite{Albash:2011nq} about the saturation of the entanglement entropy at large $\ell$ at $T=0$. The multivaluedness feeds nicely into the swallowtail structure, as we saw in the previous section by looking directly at $f(z)$.

\section{Concluding Remarks}
Since we have carefully unpacked  and discussed  our results during our presentation of  them, we will be brief in this section. 

We have presented a study, using holography, of the entanglement entropy of a certain type of strongly coupled superconductor.  Since the background is fully backÐreacted and highly stable (in the sense outlined in footnote 3), we can be confident that the results are robust.

This is the first such study of its type, and the results may well be of interest beyond the confines of holography, since it is of interest in the condensed matter physics community to use entanglement entropy as a probe of new physics of experimental relevance. Indeed, we have found that the entanglement entropy is a very sharp probe of the physics at the transition temperature $T_c$, the ground state of the system at $T=0$, and also at intermediate temperatures, where, in mapping out the full temperature range, we identified a novel\footnote{Note that while jumps in the entanglement entropy for gravity duals have been observed elsewhere in the literature~(see {\it e.g. } refs.~\cite{Nishioka:2006gr,Klebanov:2007ws,Pakman:2008ui}), our cases here (and in ref.~\cite{Albash:2010mv}) are crucially  qualitatively different from those cases since our kink in the entropy function does not result from a change in topology in the relevant minimal surfaces, but instead  arise from a multivaluedness in the entropy  from contributions of minimal surfaces of the {\it same} topology.}  transition  in the entropy. Some of the novel physics (arising from multiple extremal solutions for the entropy at a given point in parameter space) recalled phenomena observed in our earlier  studies of the  time evolution of entanglement entropy~\cite{Albash:2010mv}. The origins of the transition from a field theory perspective would be very interesting to study, and we leave that for future work. It would also be of value to study the  fate  of the physics away from the strict large $N$ limit  we have been working in here.

The study presented here also served as another holographic example of the behaviour of entanglement entropy along an RG flow, which we studied in ref.~\cite{Albash:2011nq}, and in fact we were able to confirm some more of our predictions from that paper using phenomena observed here. 

We expect that this is just the beginning of a series of fruitful investigations of this type, shedding more light on a wide variety of strongly coupled quantum phenomena of interest.

\section*{Acknowledgements}
TA and CVJ would like to thank the  US Department of Energy for support under grant DE-FG03-84ER-40168, and Nikolay Bobev and Arnab Kundu for conversations and correspondence.   We thank Jerome Gauntlett for helpful comments on an earlier version of this manuscript that led us to clarify some remarks concerning stability.  TA is also supported by the USC Dornsife College of Letters, Arts and Sciences.

\providecommand{\href}[2]{#2}\begingroup\raggedright\endgroup


\end{document}